\documentclass[10pt,a4paper,twoside]{aa}

\usepackage{psfig,latexsym,longtable}
\begin{document} 


\input{psfig.sty}
\def\sun{\hbox{$\odot$~}}
\def\deg{\hbox{$^\circ$}}
\def\kms{\,km\,s$^{-1}$}
\def\m{$^{\rm m}$}
\def\si{$\sim$}
\def\di{$\div$}
\def\av{A$_{\rm V}$ }
\def\msol{~M$_\odot$ }
\def\msolr{~M$_\odot$~yr$^{-1}$ }
\def\micron{\,$\mu$m}
\def\hi{H\,{\sc i} } 
\def\marc{mag~arcsec$^{-2}$}
\def\gae{\mathrel{>\kern-1.0em\lower0.9ex\hbox{$\sim$}}}
\def\ecs{ergs cm$^{-2}$ s$^{-1} \ $}
\def\es{ergs s$^{-1} \ $}
\def\cts{counts s$^{-1} \ $}
\def\esz{ergs s$^{-1}$ Hz$^{-1} \ $}
\def\kms{km s$^{-1}$}
\def\kcssk{$keV\ cm^{-2}\ s^{-1}\ sr^{-1}\ keV^{-1}\ $}
\def\ni {\noindent}  
\def\Msun{$M_{\odot}\ $}
\def\lae{\mathrel{<\kern-1.0em\lower0.9ex\hbox{$\sim$}}}
\def\gae{\mathrel{>\kern-1.0em\lower0.9ex\hbox{$\sim$}}}

\title{XMM-{\it Newton} observations of ULIRGs I: 
A Compton-thick AGN in IRAS19254-7245}

\author{V.\,Braito\inst{1,2}, A.\,Franceschini\inst{2}, R.\,Della Ceca\inst{3}, 
 P .\,Severgnini\inst{3}, L.\,Bassani\inst{4}, M.\,Cappi\inst{4}, G.\,Malaguti\inst{4}, 
G.G.C.\,Palumbo\inst{5}, M.\,Persic\inst{6}, G.\,Risaliti\inst{7,8} and  
 M.\,Salvati\inst{7}}
\institute{ 
  INAF $-$ Osservatorio Astronomico di Padova, Vicolo dell'Osservatorio  5, 35122 Padova, Italy. 
\and  
  Dipartimento di Astronomia, Universit\`a di  Padova, Vicolo dell'Osservatorio  2, 35122 Padova, Italy. 
\and  
INAF $-$ Osservatorio Astronomico di Brera, Via Brera 28, 20121 Milano, Italy. 
\and 
IASF $-$ CNR, Sezione di Bologna, Via Gobetti 101, 40129 Bologna, Italy.
\and 
Dipartimento di Astronomia, Universit\`a di Bologna,  Via Ranzani 1, 40127 Bologna, Italy. 
\and 
INAF $-$ Osservatorio Astronomico di Trieste, Via G. B. Tiepolo 11, 34131 Trieste,  Italy 
\and 
INAF $-$ Osservatorio Astrofisico di Arcetri, Largo E. Fermi 5, 50125 Firenze, Italy. 
\and
Harvard-Smithsonian Center for Astrophysics, 60 Garden Street, Cambridge, MA, 02138, USA.
} 
\date{Received:XX; Accepted: XXX}

\authorrunning{Braito et al.}
\titlerunning{XMM-{\it Newton} observations of IRAS~19254-7245.}

\abstract{
We present   the XMM-{\it Newton} observation of the  merging system IRAS~19254-7245, also known
as {\it The Superantennae}, whose southern nucleus is classified as a Seyfert 2 galaxy.   The XMM-{\it
Newton}  data have allowed us to perform a detailed X-ray imaging and  spectral analysis of this system.  We
clearly detect, for the first time  in this system, a strong ($EW \sim 1.4$ keV)  Fe emission line at $6.49\pm 0.1$ keV (rest-frame). The X-ray
spectrum requires a soft thermal component ($kT\sim0.9$ keV; $L_{0.5-2}\sim 
4\times10^{41}$ erg s$^{-1}$), likely associated  with the starburst, and a hard power-law continuum above 2 keV
(observed $L_{2-10}\sim 4\times10^{42}$ erg s$^{-1}$ ).  We confirm the flatness of this
latter component,  already noted in previous ASCA data. This
flatness, together with the  detection of the strong Fe-K${\alpha}$ line and other broad band indicators,
suggest the presence of a Compton-thick AGN with intrinsic luminosity $\gae 10^{44}$ erg
s$^{-1}$. We show that a Compton-thick model can perfectly   reproduce the X-ray  spectral properties of
this object.


\keywords{Galaxies: active,
 Galaxies: starburst, 
 X-rays: galaxies,
Galaxies: individual: IRAS~19254-7245}}
   
\maketitle

\section{Introduction}

The Ultra Luminous Infrared Galaxies (hereafter ULIRGs)   are  an enigmatic
class of sources  which emit most of their energy    in the far-IR (FIR)
domain,  with  luminosities in excess to 10$^{12}$ L$_\odot$,
(i.e. comparable to  QSO luminosities). While it is now clear that the huge
observed FIR luminosity is due to the presence of large amounts of dust
reprocessing the ``primary" optical and UV emission  from the central source,
the nature of the latter is still debated.  It is now widely accepted that
both starburst (SB) and  AGN activity may  be responsible for the observed
luminosity, but their relative contributions are still  unconstrained (in
some cases even the presence of an AGN is unclear).

Hard X--ray  ($E>$2 keV) observations, less affected by the photoelectric
absorption, are a fundamental tool not only to investigate the presence of an
hidden AGN, but also  to study its physical properties and to estimate its
contribution to the high observed FIR emission. Indeed some of the ULIRGs,
classified as pure SBs based on optical and IR spectroscopy, 
  show spectral  properties typical of obscured AGNs when observed in hard X-rays (i.e., NGC 6240: \cite{iwa99}).

To shed light on this topic we are carrying out a mini-survey with  XMM-{\it Newton} of
10 nearby  ($z < 0.2$) ULIRGs (see \cite{bra}) for which  high-quality mid-IR and optical
spectroscopic data are available  (\cite{gen};  \cite{lutz99}; \cite{veille}). XMM-{\it Newton}, thanks to its wide
energy band  (0.2--10 keV)  and high sensitivity (as compared to  
previous missions such as  ASCA  and  ROSAT)  allows for the first time the 
investigation of  the X-ray spectral properties of this class of sources, which are  usually  X-ray faint
(\cite{risa}).  In this paper we discuss  the 0.2--10 keV  XMM-{\it Newton} spectrum of  
IRAS~19254-7245 ({\it The Superantennae}), a well known ULIRG  
($L_\mathrm{{FIR}}=1.1\times 10^{12} L_\odot$) located at redshift of
$z=0.062$. In particular, we will argue that IRAS~19254-7245 harbors a
high-luminosity ($L\gae 10^{44}$ erg s$^-1$),
Compton-thick AGN. In this letter we assume  $H_0= 50$ km s$^{-1}$ Mpc$^{-1}$
and $q_0=0.5$.

 \section{IRAS~19254-7245:  Main Properties}

Like most  ULIRGs studied so far, IRAS~19254-7245 is a merger
system.  It shows giant tails extending out to a distance of 350 kpc, triggered 
by the   merger of two gas-rich galaxies with nuclei
  $\sim$ 9 arcsec (with $H_0= 50$ km s$^{-1}$ Mpc$^{-1}$ this corresponds to 
  $\sim 15$ kpc ) apart from each other.  In the southern nucleus, which is the
dominant source at different wavelengths,   mid-IR spectroscopy reveals
the presence of an AGN (\cite{lutz99}),  classified as a Seyfert 2 
from  optical spectroscopy (\cite{mirab};  \cite{vanz}). \\

IRAS~19254-7245 was previously observed in  hard X-rays with ASCA (\cite{ima};
\cite{pappa}).   The ASCA data  were consistent  with two possible power-law
(PL) models: an unabsorbed  very flat model ($\Gamma \sim 1$) model or  an
absorbed ($N_\mathrm{H} \sim 10^{22}$cm$^{-2}$) and steeper  ($\Gamma \simeq$ 1.7)
model. No Fe-K$\alpha$ lines in the range 6.4--7 keV  were  clearly detected,
but the upper limit on the EW of such lines was $\sim$ 1-2 keV.  The statistic
of the ASCA spectrum prevented those authors from distinguishing between a
Compton-thick (i.e.  reflection-dominated) and a  Compton-thin (i.e.
transmission-dominated) source.  Moreover, given the  ASCA's modest angular
resolution  ($\sim 2'$), it was not
possible to pin-point  the optical counterpart of the bulk of the X-ray emission.

\section{Data Reduction }

IRAS~19254-7245 was observed by XMM-{\it Newton} in March 2001 with
the   EPIC (European Photon Imaging Camera: \cite{stru} and \cite{Turner}) cameras operating in full-frame mode and the thin filter applied. Data have been cleaned and processed using
the Science Analysis Software (SAS 5.3) and analyzed using standard software
packages (FTOOLS 5.0, XSPEC 11.0). The latest calibration files  released by the EPIC team have been used.  Event files produced  from
the pipeline have been filtered from high-background time intervals and  only
events corresponding to pattern 0-12 for MOS
   and pattern 0-4 for PN have been used  (see the 
     XMM-{\it Newton} Users' Handbook \cite{Ehle2001}). The net exposure time,  after data cleaning, are $\sim 14.80$ ksec, $\sim
18.38$ ksec,  and $\sim 18.39$ ksec for PN, MOS1, and MOS2  respectively. \\

IRAS~19254-7245 is detected with a S/N ratio greater than $\sim$20  in all the EPIC cameras. In
Fig.~1 we show the  3--10 keV  contour plots from the MOS2 image overlaid on the
DSS2 optical  image of IRAS~19254-7245. The hard X-ray emission comes mostly  from the  southern Seyfert 2  nucleus 
\footnote{The soft (E$<2$ keV) X-ray emission, too, comes mostly  from the southern  nucleus.   The soft X-ray emission
appears to be extended and it includes  the whole system. A detailed spectral and imaging
analysis of this  component will be reported elsewhere (Franceschini et al.  2002).
}.
Since in this paper we want to address the X-ray spectral properties of the  nuclear X-ray
emission, X-ray counts have been extracted  from  a circular region of 15 arcsec radius
\footnote{The PSF  of XMM-{\it Newton} does not allow us to extract the spectra from a  smaller
region.  
},
positionally coincident  with the core of the hard X-ray emission.    Background counts have
been extracted from a  closeby source-free   circular region. 
We have then generated our own
response matrices (that include the correction for the effective area) using the SAS tasks {\it arfgen} and {\it rmfgen}.
 
 In order to improve the statistics, the MOS1 and MOS2 data have been combined
together and the  combined MOS and PN spectra have been   fitted
simultaneously  keeping the relative normalizations free.  All the models
discussed in this paper have been filtered through the Galactic absorbing column
density along the line of sight to IRAS~19254-7245 ($N_\mathrm{H} = 5.95\times 10^{20} $cm$^{-2}$).
 Unless otherwise stated, in what follows errors will be given at the 90\%
confidence level for one interesting parameter ($\Delta\chi^2$=2.71).
\begin{figure}[htb]
\parbox{10cm}{
\psfig{file=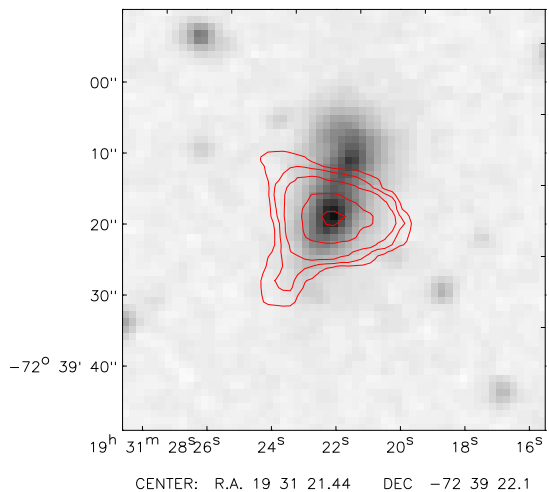,height=7.0cm,width=10cm,angle=0}}
\caption {DSS2 image ($1'\times 1'$) of the field centered  on IRAS
19254-7245. Contours of the hard (3--10 keV)  X-ray emission  have been  
overlaid on the optical image. We have used here the MOS2 data since 
this detector has the best point spread function
(see \cite{Ehle2001}). The displayed X-ray contour levels correspond to 5$\sigma$, 7$\sigma$, 10$\sigma$, 20$\sigma$, 30$\sigma$ above the background.}
\end{figure}

\section{Spectral Analysis}

Single-component models, e.g. a  thermal model
\footnote{The thermal emissions used in this paper 
are described by the MEKAL  (\cite{mewe})  model in the XSPEC package.}
 or a  PL model, are
both rejected by the data at a high confidence level.
In Fig.~2 we show the ratio between the best-fit PL model and  the
data: large excesses are clearly present in the soft energy band (suggesting  the presence of a 
soft thermal component) and a ``line like" feature is evident  between 6 and 7
keV (suggesting the presence of an Fe line).  On the other hand a single 
thermal component cannot account for the emission at $E>2 $ keV.\\ 

We then tried to fit the data with a thermal component, an absorbed PL
component plus a Gaussian line around  $6.4$ keV. The best-fit values of this
model are: $kT=0.86^{+0.15}_{-0.09}$ keV,  $\Gamma= 1.3^{+0.14}_{-0.21}$, 
$N_\mathrm{H}=3.1^{+3.1}_{-1.5}\times 10^{21}$cm$^{-2}$ (model A in Table 1
and Table~2).  The energy  position and the EW of the Gaussian line are
$E=6.48_{+0.10}^{-0.10} keV $ and  $EW =1.4_{+0.5}^{-0.5}$ keV  (source rest-frame). The energy position of this line is consistent with cold Fe-K$\alpha$
emission. It is worth noticing that the line profile appears to be  marginally
broad,  with a  width $\sigma=0.3^{+0.2}_{-0.2}$ keV. This broadening can be due
to the merging of several lines  or to a more complex modeling of the continuum
(e.g. the presence of an absorption edge modifying the underlying continuum,
see below).

\begin{table*}[tb]
\caption{ Results from the best-fit model. }

   \begin{tabular}[h]{lcc ccc  cccc }
      \hline
      \hline
     
   Model& $KT$ & Norm$^{a}$ & $\Gamma$ &  $N_{\mathrm{H}}$  & Norm$^{b}$ & Norm$^{c}$  & $E_{\mathrm{k}\alpha}$ &  $EW$ &$\chi^2/dof$\\

      \hline
         & keV&  & & $10^{22}$cm$^{-2}$ & & &keV&\\
\hline
 A & 0.86$^{+0.15}_{-0.09}$	 & 1.23$^{+0.30}_{-0.31}$ & 1.30$^{+0.14}_{-0.21}$ & 0.31$^{+0.31}_{-0.15}$ &
 2.63$^{+1.09}_{-0.71}$&/&
  6.48$^{+0.10}_{-0.10}$  &1.4$\pm 0.5$&50.9/57\\

  B & 0.85$^{+0.14}_{-0.09}$ &1.26$^{+0.37}_{-0.39}$ &1.84$^{+0.60}_{-0.55}$   & 0.47$^{+0.52}_{-0.25}$ & 
  3.71$^{+2.40}_{-1.40}$ & 46.4$^{+181.2}_{-43.5}$&6.49$^{+0.09}_{-0.10}$  &2.0$\pm 0.6$&48.1/56\\

 \hline
  \end{tabular}

$^{a}$ in units of $10^{-19} / (4 \pi (D^{2}) \int n_e n_H dV$ , where D is 
          the  distance to the source (cm), n$_e$ is the electron 
          density (cm$^{-3}$), and n$_\mathrm{H}$ is the hydrogen density (cm$^{-3}$) at 1 keV.
 \\
$^{b}$ in units of  10$^{-5}$ photons keV$^{-1}$ cm$^{-2}$ at 1 keV; in model B it refers to the scattered component. \\
$^{c}$  in units of  10$^{-5}$ photons keV$^{-1}$ cm$^{-2}$ at 1 keV; in model B it refers to the reflected component.
\end{table*}

\begin{table*}[tb]
\caption{ Observed X-ray Fluxes and X-ray luminosities }
  
   \begin{tabular}[h]{lc ccc cccc }
      \hline
      \hline
      
   Model&BAND & &  \multicolumn{3}{c}{Flux}  & &\multicolumn{2}{c}{Luminosity$^b$} \\
   &&& \multicolumn{3}{c}{10$^{-14}$ erg cm$^{-2}$ s$^{-1}$}& &\multicolumn{2}{c}{10$^{41}$ erg s$^{-1}$} \\
   \cline{4-6}\cline{8-9}
   
&  & &TOTAL & Starburst& AGN & &Starburst & AGN\\
\hline\hline
A& 0.5--2 keV&& 5.30 &2.13 & 3.16 & &4.4$^a$ &    9.7$^b$\\ 

 & 2--10 keV&&  23.0  &0.1 & 22.9 & &0.30$^a$ &37.7$^b$\\
 
 B& 0.5--2 keV&& 5.31  &2.10 & 3.21 && 4.3$^a$&13.6$^b$\\ 

 & 2--10 keV   &&23.5&0.1& 23.4 & &0.28$^a$ &38.7$^b$\\
 \hline
  \end{tabular}
  
$^{a}$  X-ray luminosity de-absorbed from intrinsic and galactic $N_\mathrm{H}$.  \\
$^{b}$  X-ray luminosity de-absorbed from  galactic $N_\mathrm{H}$.  \\

\end{table*}

The flat photon index, $\Gamma\sim 1.3$, measured for the hard X-ray continuum, and the presence of a Fe-K$\alpha$ emission are both typical of
high-mass X-ray binaries (HMXB) spectra (\cite{white}). However, the equivalent width
of the Fe-K complex arising from HMXBs,  either observed directly in HMXBs (typically
EW $\sim$ 0.3 keV: see White et al. 1983) or inferred for  galaxies with X-ray emission
dominated by HMXBs (\cite{persic}), is not consistent  with the huge value observed
here. The flat hard  continuum and the prominent Fe-K$\alpha$ emission line, may alternatively suggest that IRAS~19254--7245
hosts  a Compton-thick AGN (e.g. \cite {maiol}). In this perspective   we have   tried a composite model which accounts for the SB
emission  and for the  ``putative'' Compton-thick AGN emission. The SB emission has been
modeled with a soft thermal component. The  AGN emission has been described by a pure Compton-reflected  continuum (model PEXRAV
\footnote{Since we
lack  any information about the emission at energy above 10 keV  we  
kept the inclination, the abundances  and the energy cutoff frozen to
standard values of $\sim 60^\circ$, 1 and 100 keV; only the intrinsic
PL photon index and the reflection normalization  were  allowed to
vary in the fitting procedure.}
in XSPEC, \cite{pexra}) combined  with an absorbed PL
model and a Gaussian line  at $\sim 6.4$ keV.  In this model the absorbed
PL component represents the  scattered emission of the  central
source, hence its  photon index is tied to the one in the
PEXRAV model. The results are reported in Tables~1 and ~2 (model B) and shown in Fig.~3. 
We note that the best-fit photon index is
$\sim$1.8, a value  very similar to that observed in unobscured AGN (\cite{george}). 

As described  above, the baseline model for the  hard X-ray continuum of IRAS~19254--7245 comprises a
reflected plus a scattered  component.  If this latter component is supposed to arise   from a  ``warm
mirror'', then  one would also expect to observe emission lines from   highly ionized Fe.  We have
estimated  the upper limits on the EW  of  the He-like (6.7 keV) and the H-like (6.97 keV) Fe emission lines
to be $\sim$ 1.8 keV and  $\sim$2 keV,  respectively.  A second narrow gaussian line  could probably be
 accommodated within the data uncertainties. These upper limits are consistent with that expected
from the scattered component, but the present  statistics does not allow us to be conclusive about this
point.

In summary, the broad-band X-ray emission from IRAS~19254--7245 requires  a 
thermal emission ($ kT \sim 0.85$ keV) plus a hard  continuum.  The spectral
parameters of the thermal emission are  not a strong function of    the
detailed modeling of the hard X-ray continuum; this component (which dominates
at $E\lae 1$ keV)  is likely
associated with the starburst and, compared to the hard one, appears to be
spatially extended.  A detailed analysis of this emission will be presented  in
a forthcoming paper (Franceschini et al. 2002). The hard X-ray  continuum can
be described by two models: a) a Compton-thin (Table~1 model A) AGN having a
very flat photon index or  b) a Compton-thick (Table~1 model B) AGN having an
intrinsic photon index of   $\sim$1.8. Both  models require a strong Fe line at
$\sim$6.49 keV (rest-frame). Since  these two competing models give similar $\chi^2/\nu$ in
the  next section we will try to discriminate between them.

\begin{figure}[htb] 
\parbox{10cm}{\psfig{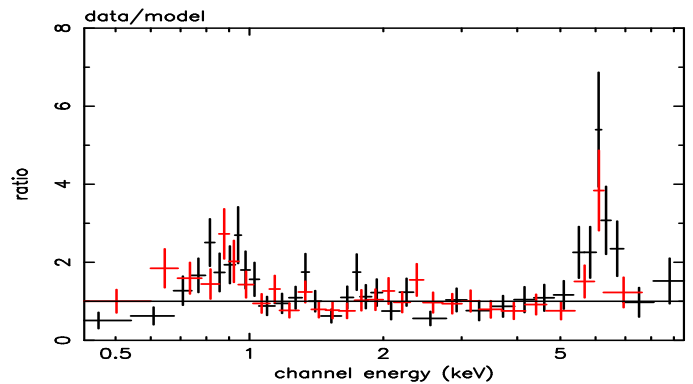}}  
\caption[]{Ratio between the PN (red) and MOS (black) data and the model  when  the spectrum of  IRAS~19254--7245  is fitted with a single power law model; strong residuals are present at
energy below 1 keV and above 5 keV.}
\end{figure}

\begin{figure}[ht]
\parbox{10cm}{ \psfig{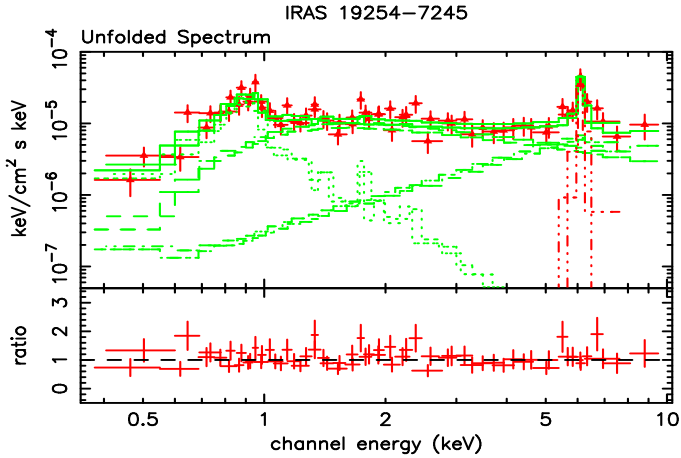}
      
}

\caption[]{IRAS~19254-7245 (Model B). Upper panel: unfolded X-ray spectrum. Lower
panel: ratio between the data (MOS+PN) and the best-fit
model.} \end{figure}

\section{Discussion} 

A viable interpretation of the observed flat ($\Gamma \sim 1.3$) hard X-ray continuum and
prominent ($EW \sim$ 1.4 keV) Fe-K$\alpha$ emission line involves a reprocessing of the primary spectrum,  i.e.
the presence of a Compton-thick source.
Further indications of the Compton-thick nature of IRAS~19254--7245  
come  from  observations at other wavelengths.  One of the
broad-band diagnostics  largely  used in the literature to estimate the
strength  of the nuclear source and the Compton thickness is the
ratio between the strength of the [O$_{III}$]$\lambda$5007 emission
line and the 2--10 keV  flux. This ratio -once that the 
[O$_{III}$] emission is corrected  for the extinction using the
Balmer decrement (\cite{bas})-  combined  with the EW of the
measured Fe emission line  can be used to evaluate the Compton
thickness of the source.

Using recent  [O$_{III}$] estimates (for the southern nucleus; see Berta et al. 2002) and  the 2-10 keV observed flux (see Table~2) we obtain
F(2--10 keV)/F[O$_{III}$] $\sim 0.09$. This  value, combined with the measured $EW_{Fe-K\alpha} \sim 1-2$ keV of the
Fe line,  clearly  locates   IRAS 19254--7245 in the region populated by 
Compton-thick  sources in  Fig.~1 of \cite{bas}. 

The energy position of the line ($6.49^{+0.09}_{-0.10}$ keV) is consistent  with cold
(i.e. Fe less ionized than Fe XVII) Fe K$\alpha$ emission but not with highly ionized
(He-like or H-like) iron.  Therefore we can exclude an origin of the line from 
optically thin gas (the so called ``warm mirror") placed outside the torus (see e.g. 
\cite{matt96}), leaving the production from a cold medium as the only  possibility.  In
this context the line could be produced by reflection from the inner surface of the
circumnuclear absorbing material (\cite{ghise}) or by transmission  through the torus itself
(\cite{lea}).  Moreover the X-ray spectrum is well described by a reflection component,
and the  equivalent width of the detected line (with respect to the reflected 
component) is 2 keV: this is in good agreement with the expectation  from a
reflection dominated source (see \cite{matt96}).

If the Compton-thick hypothesis is correct, then  the observed  luminosity  (see Table~2) could
be just few percent of the {\it intrinsic} one.  Since the best-fit luminosity  of
the scattered  component is $L^{\mathrm{SC}}_{(2-10)} \sim 10^{42}$ erg s$^{-1}$,
then the intrinsic luminosity is estimated to be $L^{\mathrm{INT}}_{(2-10)}= L^{\mathrm{SC}}_{(2-10)}\tau^{-1}(\Omega/2\pi)^{-1}$. Assuming   the NGC1068 values
(\cite{iwa97}) for the mirror optical depth ($\tau\sim 10^{-3}$) and the
subtended angle ($\Omega/2\pi=0.25$),   we obtain an  intrinsic luminosity
for the AGN harbored in IRAS~19254-7245  of $L^{\mathrm{AGN}}_{(2-10)}\sim 10^{45}$erg s$^{-1}$.\\

The derived intrinsic luminosity of the AGN present in IRAS~19254-7245
suggests that the AGN activity contributes a significant
fraction of the FIR luminosity.  Indeed mid-IR observations
suggest that the AGN contribution to the FIR emission could be 
40\% (see Berta et al. 2002).  We have then
computed   log( $L_{\mathrm{FIR}}/L_\mathrm{2-10}$) for the AGN
component. We found that this  ratio is $\sim 3 $,  in agreement with the Compton-thick  scenario (\cite{mul}).  

In Fig.~4 we compare the spectral  energy distribution (SED) of
IRAS~19254-7245  with the SED of NGC6240 (\cite{iwa2001}).  The latter ULIRG is
known to harbor a heavily-obscured,  high-luminosity AGN.  In this respect it
is worth emphasizing the similarities of the two SEDs from the IR--to--X-ray
band: in particular, both sources show a strong Fe emission line superimposed
on a strongly  absorbed continuum.  In more detail NGC6240 and IRAS~19254-7245
show a similar emission in the X-ray  band, while they differ in the
optical/FIR band (IRAS~19254-7245 being  more luminous then  NGC6240). This
could indicate that the AGN in IRAS~19254-7245 could be intrinsically more
powerful and/or more obscured in the X-ray energy band.  Alternatively, the  SB
activity  may be stronger in  IRAS~19254-7245 than in NGC6240. Finally the
shapes  of these SEDs are  similar to the SED of the cosmic energy density
spectrum (see \cite{lehm} and reference therein). In both ULIRGs there is
evidence of comparable contributions by obscured SB and AGN activities.

\begin{figure}[bht]
\parbox{10cm}{ \psfig{file=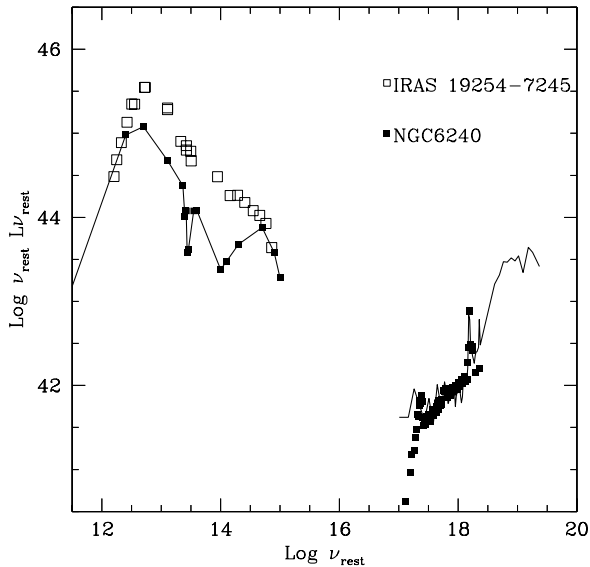,height=8.0cm,width=8.0cm,angle=0}}

\caption[]{Comparison between the SED of the Southern nucleus of IRAS~19254-7245 (open squares) and NGC6240
(filled squares and solid line). The data of the SED of NGC 6240  are from  \cite{iwa2001}. The data of IRAS~19254-7245 are from Berta  et al. 2002.}

\end{figure}

\section{Conclusions} 

We have presented the XMM-{\it Newton} observations of
IRAS~19254-7245. The main results  are  hereby summarized. 
\begin{enumerate} 
 \item The hard  X-ray emission  originates from the
southern nucleus of IRAS~19254-7245, which is  optically classified
as a Seyfert 2 galaxy. 
\item We detect, for the first time in this system, a strong ($EW \sim 1.4-2.0$ keV)
emission line at $6.49\pm 0.10$ keV (rest-frame), which is 
consistent with a cold Fe-$K{\alpha}$  line.
  \item The broad-band X-ray spectra requires a soft thermal
  component, which accounts for the starburst emission,  
  and a flat ($\Gamma \sim 1.3$) continuum most likely associated with a heavily absorbed  AGN. 
The hard continuum and the  detected  strong Fe-K$\alpha$
 emission line are highly  indicative of a Compton-thick source. Indeed, from a
statistical and a physical point of view, the X-ray  spectral
properties of this object can be well explained with a Compton
thick  model. In this scenario, the intrinsic 2--10 keV 
luminosity of  IRAS~19254-7245 could be greater than   $10 ^{44}$
erg s$^{-1}$, i.e. attaining the QSO luminosity regime.  
\end{enumerate}
\begin{acknowledgements}  This work has received financial support from
ASI (I/R/037/01) under the project ``Cosmologia Osservativa con
XMM-Newton" and  from the Italian Ministry of University and
Scientific and Technological Research (MURST) through grants Cofin
$00-02-004$. PS acknowledges partial financial support by the
Italian {\it Consorzio  Nazionale per l'Astronomia e l'Astrofisica}
(CNAA).  We thanks K. Iwasawa to have provided us the SED of NGC6240 in a tabular form.

\end{acknowledgements} 


\begin{thebibliography}{}

\bibitem[Bassani et al. 1999]{bas}
Bassani L., Dadina M., Maiolino R. et al. 1999, ApJS 121, 473
\bibitem[Berta et al 2002]{berta}
Berta, S., Fritz J., Franceschini, A., Bressan, A. \& Pernechele,  C. A\&A submitted  
\bibitem[Braito et al. 2002]{bra}
Braito, V., Franceschini, A., Della Ceca, R. et al. 2002  ``New Visions of the
 X-ray Universe in the XMM-Newton and Chandra Era", ESA-SP, ed. F. Jansen 
 (astro-ph/0202352)
\bibitem[Ehle et al. 2001]{Ehle2001}
Ehle, M., Breitfellner, M., Dahlem, M.  et~al.\ 2001, XMM-Newton Users' Handbook
\bibitem[Franceschini et al. 2002]{franc}
Franceschini, A., Braito, V., Della Ceca, R.  et al. in preparation
\bibitem[Genzel et al. 1998]{gen}
Genzel R., Lutz D., Sturm E. et al. 1998, ApJ 498, 579
\bibitem[George et al. 2000]{george}
George, I.~M.,  Turner, T.~J.,  Yaqoob, T. et al. 2000, ApJ, 531, 52
\bibitem[Ghisellini et al. 1994]{ghise}
Ghisellini, G., Haardt, F. \& Matt, G. 1994 , MNRAS, 267, 743
\bibitem[Imanishi et al. 1999]{ima}
Imanishi, M. \& Ueno, S. 1999, ApJ, 527, 709
\bibitem[Iwasawa et al. 1997]{iwa97} 
Iwasawa, K., Fabian, A.~C. \& Matt, G., 1997, MNRAS, 289, 443
\bibitem[Iwasawa 1999]{iwa99}
Iwasawa, K. 1999, MNRAS, 302, 96
\bibitem[Iwasawa et al. 2001]{iwa2001}
Iwasawa, K., Matt, G., Guainazzi, M. \& Fabian, A.~C., 2001, MNRAS, 326, 894
\bibitem[Leahy   \& Creighton  1993]{lea}
Leahy, D.~A.  \& Creighton, J., 1993, MNRAS, 263, 314
 \bibitem[Lehmann et al.  2001]{lehm}
 Lehmann, I., Hasinger G., Murray, S.~S. \& Schmidt M.,  Proceedings for X-rays at Sharp Focus Chandra Science Symposium, held in St. Paul, MN (2001), eds. E~M. Schlegel and S.Vrtilek, 2001 heus confE, 10
\bibitem[Lutz et al. 1999]{lutz99}
Lutz, D. and Veilleux, S.  \& Genzel, R.  1999, ApJ, 517L, 13.
\bibitem[Magdziarz \& Zdziarski 1995]{pexra}
Magdziarz, P. \& Zdziarski,  A.~A. 1995, MNRAS, 273, 837
\bibitem[Maiolino et al. 1998]{maiol}
Maiolino, R., Salvati, M., Bassani, L. et al. 1998,  A\&A, 338, 781.
\bibitem[Matt et al. 1996]{matt96}
Matt, G., Brandt, W.~N. \& Fabian, A.~C., 1996, MNRAS, 280, 823
\bibitem[Mewe et al. 1985]{mewe}
Mewe R., Gronenschild E.~H.~B.~M. \& van den Oord G.~H.~J., 1985, A\&AS 62, 197 
\bibitem[Mirabel et al. 1991]{mirab}   
Mirabel, I.~F., Lutz, D. \& Maza, J. 1991, A\&A, 243, 367
\bibitem[Mulchaey et al. 1994]{mul}   
 Mulchaey, J.~S. Koratkar, A., Ward, M.~J. et al., 1994, ApJ, 436, 586
\bibitem[Pappa et al. 2000]{pappa}  
Pappa, A., Georgantopoulos, I. \& Stewart, G.~C. 2000, MNRAS, 314, 589
\bibitem[Persic \& Rephaeli 2002]{persic}
Persic, M. \& Rephaeli, Y. 2002, A\&A, 382, 843 
\bibitem[Risaliti et al. 2000]{risa}
Risaliti, G., Gilli, R., Maiolino, R. \& Salvati, M. 2000, A\&A, 357, 13
\bibitem[Str\"uder et al. 2001]{stru}
Str\"uder, L., Briel, U., Dannerl, K., et al. 2001, A\&A, 365, L18
\bibitem[Turner et al. 2001]{Turner}
Turner, M. J. L., Abbey, A., Arnaud, M., et al. 2001, A\&A, 365, L27
\bibitem[Vanzi et al. 2002]{vanz}
Vanzi, L., Bagnulo, S., Le Floc'h,  E. et al., 2002, A\&A, 386, 464
\bibitem[Veilleux et al. 1999]{veille}
Veilleux S., Kim  D.-C.  \& Sanders, D.~B. 1999, ApJ 522, 139 
\bibitem[White et al. 1983]{white}
White, N.~E., Swank, J.~H. \& Holt, S.~S.,1983, ApJ, 270, 711
\end{thebibliography}
\end{document}